\documentclass[12pt,twoside]{mitthesis}
\usepackage{lgrind}
\usepackage{amsmath, amssymb, amsthm, amscd, verbatim}
\usepackage{tikz}
\usepackage{enumerate}
\usetikzlibrary{matrix,arrows}
\usepackage{pgfplots}

\usepackage{setspace}
\usepackage{graphicx}
\usepackage{cmap}
\usepackage{hhline}
\theoremstyle{plain}
\newtheorem{theorem}{Theorem}

\newtheorem{lemma}[theorem]{Lemma}
\newtheorem{corollary}[theorem]{Corollary}

\theoremstyle{definition}
\newtheorem{definition}[theorem]{Definition}

\theoremstyle{remark}

\begin{document}

%
%
%
%
%
%
%
%
%
%
%

\title{Finding the Optimal Demodulator Under\\ Implementation Constraints}

\author{Qian Yu}
       \prevdegrees{B.S. Physics and EECS, \Mit, 2014}
\department{Department of Electrical Engineering and Computer Science}

\degree{Master of Engineering in Electrical Engineering and Computer Science}

\degreemonth{June}
\degreeyear{2015}
\thesisdate{May 22, 2015}



\supervisor{Jason Pearce}{Director, IC Design, Maxim Integrated}

\supervisor{Muriel Medard}{Cecil H. Green Professor of Electrical Engineering, EECS, MIT}

\chairman{Albert R. Meyer}{Chairman, Masters of Engineering Thesis Committee}

\maketitle



\cleardoublepage
\setcounter{savepage}{\thepage}
\begin{abstractpage}
%
%
%
The common approach of designing a communication device is to maximize a well-defined objective function, e.g., the channel capacity and the cut-off rate. We propose easy-to-implement solutions for Gaussian channels that approximate the optimal results for these maximization problems. Three topics are addressed.

First, we consider the case where the channel output is quantized, and we find the quantization thresholds that maximize the mutual information. The approximation derived from the asymptotic solution has a negligible loss on the entire range of SNR when 2-PAM modulation is used, and its quantization thresholds linearly depend on the standard deviation of noise. We also derive a simple estimator of the relative capacity loss due to quantization, based on the high-rate limit.

Then we consider the integer constraint on the decoding metric, and maximize the mismatched channel capacity. We study the asymptotic solution of the optimal metric assignment and show that the same approximation we derived in the matched decoding case still holds for the mismatched decoder. 

Finally, we consider the demodulation problem for 8PSK bit-interleaved coded modulation(BICM). We derive the approximated optimal demodulation metrics that maximize the general cut-off rate or the mismatched capacity using max-log approximation . The error rate performances of the two metrics' assignments are compared, based on Reed-Solomon-Viterbi(RSV) code, and the mismatched capacity metric turns out to be better. The proposed approximation can be computed using an efficient firmware algorithm, and improves the system performance of commercial chips.
 

\end{abstractpage}


\cleardoublepage

\section*{Acknowledgments}

I am grateful to my direct supervisor at Maxim Integrated, Jason Pearce, and his group: Anish Shah, Siddharth Gupta, and Milton Cheung. I received constant support from them on design and debugging during the 9 months I spent at Maxim. Having them as coworkers made my work truly enjoyable. Many thanks to Cheng-Wei Pei for helping me match with this amazing group.

I would like to thank my thesis advisor Muriel M\'{e}dard, who gracefully agreed to supervise my work, and provided me with much useful guidance. I appreciate all the additional effort she made and her valuable advice on thesis-related work.

I would also like to thank Robert A. Irwin from the MIT Writing and Communication Center, who provided many extremely valuable tips on English writing, in addition to his advice on revising my thesis. I also owe thanks to my fianc\'{e}e Siyao Xu, who not only helped with revising my paper but also provided unwavering mental support throughout these years.

Finally, many thanks to my family and all my friends, who kindly backed me up during my hard times. I would not have completed this project without all their help.


\pagestyle{plain}
\tableofcontents
\newpage
\listoffigures
\newpage
\listoftables

\chapter{Introduction}


A common goal of designing a communication system is to minimize the chances of getting incorrect messages, given the transmission rate and other implementation constraints. However, the actual error rate non-trivially depends on the configuration of the entire signal chain and the channel condition; thus it cannot be used as the cost metric when optimizing the design of a simple block.


Various objective functions have been proposed (e.g., the channel capacity and the cut-off rate) to be used as alternative metrics in the optimization problem. Numerical solutions were derived in many past studies based on these metrics.

The objective of this thesis is to solve the above optimization problem while taking the implementation cost into account. We derive results that are simple to implement, yet approximate well the optimal solutions. We then compare the error rate between the solutions based on different metrics. We also show that the optimized solution derived in this thesis can be implemented on commercial chips with an improvement of system performance. 

\section{Overview of Digital Communication Systems}

\begin{figure}
	\centering
	\includegraphics[width=120mm]{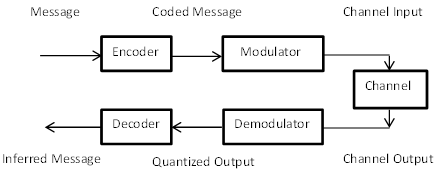}
	\caption{The basic structure of a digital communication system.}
	\label{arm:fig1}
\end{figure}

The basic structure of a modern digital communication system consists of two functions: error correction and analog-digital conversion (see Fig.\ref{arm:fig1}).

The physical channel that delivers the information is commonly modeled as a randomized analog function, which corrupts the transmitted signal and creates error. Since the error correction codes operate in the digital domain, it is necessary to have the analog-digital conversion.

The modulator maps strings of bits from the coded message to values in a fixed set of constellation points; then the demodulator quantizes the channel output back into discrete values.

The non-ideal conversion between analog and digital signals lowers the system performance, so it is valuable to optimize these conversion blocks to achieve the best performance given implementation constraints.

The transmitter path is usually fully characterized by the standard, but the same restriction does not apply to the receiver. Hence, we want to optimize the design of the receiver, and specifically the design of the demodulator.   

\section{Background}

The optimal receiver, one that minimizes the error probability, should obey the maximum likelihood (ML) decision rule. This ideal condition is not practically achievable, due to implementation constraints. These includes the upper limit of the quantization resolution, the integer constraints on the decoding matrices, and the use of bit-interleaved coded modulation (BICM). 

There have been extensive studies on the optimal receiver design, given the above restrictions.

When the only constraint is the number of quantization regions, the mutual information is commonly used as the optimization metric. Numerical solutions of quantization thresholds were derived for 2-PAM Gaussian channels \cite{rave2009quantization}. Since deriving a numerical solution in the general case is considered difficult, several sub-optimal approaches have also been discussed \cite{chandrasekaran2011capacity,danieli2010maximum,liveris2003quantization,novak2009quantization,shahid2008distributed}. 

When the other two constraints are also applied, the problem is commonly modeled as a mismatched decoding problem. A lower bound of the highest achievable rate for mismatched decoding is discussed in \cite{hui1983fundamental,merhav1994information}, which is later referred to as the generalized mutual information (GMI). The GMI bound is equal to the mismatched capacity in some special cases\cite{ahlswede1996erasure,balakirsky1995converse}, and a generalized version of GMI was recently proved to be a tight bound \cite{somekh2014general}. Therefore, the GMI is commonly used as the objective function in optimization problems \cite{abbe2007finding,binshtok1999integer}.

The generalized cut-off rate (GCR) was introduced in \cite{biederman1981decoding} as the cut-off rate in the mismatched case. It is another commonly used objective function. Investigations of the GCR maximization problem under integer metrics constraint have been made in \cite{binshtok1999integer,salz1995decoding}.

\section{Thesis Layout}
We study the optimal demodulator implementations and their approximations for Gaussian channels, under different implementation constraints.

In chapter \ref{cp:matched}, we start with the simplest case, in which the decoder allows matched decoding, and we find the optimal quantization that maximizes the mutual information. In chapter \ref{cp:mm}, we add the integer decoding metric constraint into consideration and generalize our results to the mismatched channel capacity. In chapter \ref{cp:bt}, we investigate the optimal decoding metrics assignment for 8PSK BICM systems. Then, in chapter \ref{cp:cl}, we summarize our results and propose future work from this thesis.

\section{Preliminaries}
\label{pre}
In this thesis, we make the following assumptions: 

We consider the memoryless channels, with their inputs denoted by $X$ and their outputs denoted by $Y$.

For each channel, the output is linear to the input with gain $g$, and there is an additive Gaussian noise $W$ independent of the input. In other words,
\begin{align}
\label{chn}
Y=gX+W
\end{align}
Without loss of generality, we assume that the channels are always normalized so that the inputs and the noises have zero mean and unit variance.












\chapter{Optimization Quantizer for Matched Decoding}
\label{cp:matched}
This chapter describes the approximation solutions of the capacity maximal quantization problem. We first briefly review the algorithmic approach of obtaining the numerical solution, then develop a novel approach of obtaining low complexity approximation schemes by observing the asymptotic solutions. We also provide a simple estimator of the capacity loss due to quantization.

\section{Problem Definition}

Consider a scalar memoryless channel, with its transition density function characterized by $f_{Y|X}$. We assume that the channel input takes its value from a fixed constellation, with a given distribution $P_X$.

\begin{figure}
 	\centering
 	\includegraphics[width=110mm]{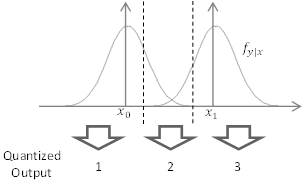}
 	\caption{An example of the analog-digital conversion.}
 	\label{arm:fig2}
 \end{figure}

The quantization process takes the channel output and converts it to an integer from a finite set: $\{1,2,...,N\}$ . We use $Z$ to denote the quantized version of the channel output (see Fig. \ref{arm:fig2}). 

In this case, $N$ represents the number of possible quantized outputs, and we consider the upper limit of $N$ as a constraint in our optimization problem.   

The optimal quantization refers to schemes $Z(Y): \mathbb{R} \rightarrow \{1,2,...,N\}$ that maximize the highest achievable information rates on such channels, which is to maximize the mutual information of the equivalent channel $I(X;Z)$, based on Shannon's paper\cite{shannon2001mathematical}:

\begin{align}
I(X;Z)= \sum_{x,z}  P_{X,Z}(x,z) \ln \frac{P_{X|Z}(x|z)}{P_{X}(x)}dy \text{.}
\end{align}

A necessary condition of the optimal solution can be derived in the following way: Given an arbitrary mutated scheme $Z'(Y)$, which only differs from the optimal solution $Z(Y)$ in an extremely small interval $[y_0,y_0+\delta]$, we calculate the change of the mutual information:
\begin{align}
\Delta I(X;Z)&\approx \sum_{X,Z}  ln ( P_{X|Z})\Delta P_{X,Z}\\
&\approx \sum_X P_X \int_{y_0}^{y_0+\delta} f(y|x) \ln \frac{P(x|z'(y))}{P(x|z(y))}dy \text{.}
\end{align}

Since the optimal solution can never be improved, the choices of $Z$ satisfy the following equation:
\begin{align}
\label{req}
Z(y)=\mathop{argmax}_{Z} \{\sum_X{P_X f(y|x)\ln(P_{X|Z})}\} \text{.}
\end{align}

This formula takes an arbitrary quantization scheme and calculates a revised scheme with mutual information improvement. Thus the numerical solution can be calculated by repeatedly executing this operation.

Although we are able to derive the optimal solution for any specific channel, sometimes it is also desirable that the quantization scheme can be adjusted when there are multiple possible channel conditions. For example, we may want to adjust the quantization thresholds for variable SNR Gaussian channels. 

The direct implementation of this adjustment requires an array of lookup tables, which is complex. More practically, approximated solutions that analytically depend on the channel parameters (e.g., SNR in the Gaussian channel case) are desired. Techniques like curve-fitting have been used in solving this problem\cite{rave2009quantization}. 

In this rest of this chapter, we aim at obtaining such simple schemes based on the asymptotic solutions of the quantization thresholds.

\section{An Upper Bound on Number of Intervals That Map to the Same Quantized Output}
\label{mulr}

To describe a quantization scheme, a common approach is to describe all the intervals on the channel output space that are mapped to each quantized output. For each quantized output $z$, we call the corresponding set of intervals the \textit{quantization intervals} of $z$, denoted by $S(z)$. 

Since (\ref{req}) does not directly rule out the possibility of an arbitrarily large number of disjoint intervals being mapped to the same quantization output, we provide an upper bound on that number, in order to simplify the implementation of numerical search algorithms and our discussion.

\begin{theorem}
	\label{tm1}
Given an optimally quantized K-PAM modulated Gaussian Channel, with the number of possible quantized outputs to be $N$, the number of disjoint intervals that map to any quantized output is upper bounded by
 $(N-1)\lfloor\frac{K-1}{2}\rfloor+1$.
\end{theorem}

The proof is contained in Appendix \ref{ap1}.

A useful corollary immediately follows for the special case of 2-PAM modulated channel:

\begin{corollary}
	\label{tm2}
	For 2-PAM Gaussian Channel, the optimal quantization always assigns single intervals to each quantized output. 
\end{corollary}

Hence in the rest of this chapter, when 2-PAM is used, we can simply treat the quantized output as a single interval on the real axis.

\section{Asymptotic Solutions to the Optimal Quantization Threshold on 2-PAM Modulation with 3 Quantization Outputs }
\label{2p3qex}

In this section, we consider 2-PAM modulated channel, with input constellation at $\pm1$, and channel output quantized into $3$ possible values.

We assume that the channel output is quantized into $3$ symmetric regions :$(-\infty,-b)$, $(-b,b)$, $(b,+\infty)$. We also assume that the channel input is distributed uniformly: $P_X(-1)=P_X(1)=\frac{1}{2}$.

We now apply (\ref{req}), and obtain the following equation:
\begin{align}
\label{2-3}
2bg=\ln(-\frac{\ln(2Q(b+g)/(Q(b+g)+Q(b-g)))}{\ln(2Q(b-g)/(Q(b+g)+Q(b-g))}) \text{.}
\end{align}
Here the $Q(x)$ is the Gaussian $Q$-function, defined as:
\begin{align}
Q(x)=\int_{x}^{+\infty} \frac{1}{\sqrt{2\pi}}e^{-\frac{k^2}{2}} dk \text{.}
\end{align}
\subsection{Large SNR Limit}

We first consider the case where the SNR is large, i.e.,  $g\gg 1$.

In order to simplify equation (\ref{2-3}), we quote a useful approximation, supported by \cite{borjesson1979simple}:
\begin{align}
\label{logq}
Q(x) \approx \frac{1}{\sqrt{2\pi}x}e^{-\frac{x^2}{2}}  \text{ \ \ \ \ \ \ \ for }x\gg1
\end{align}

Based on the above approximation, we can simplify the RHS of (\ref{2-3}):
\begin{align}\label{rhs}
\mathop{ RHS}_{g\to +\infty}=
\left\{\begin{array}{@{}ccc@{}}
\ln(\frac{(b+g)^2}{2 \ln 2}), \hfill b<g\\
\\
\ln(\frac{2bg}{ \ln 2}), \hfill b>g
\end{array}
\right.
\end{align} 

(\ref{rhs}) shows that for any fixed non-zero value of $b$, the RHS increases much slower than the LHS of (\ref{2-3}) when $g$ is large. This indicates $\lim_{g\to \infty}{b}=0$.

We use this fact and resolve (\ref{2-3}), and the first significant term of $b$ is thus:
\begin{align}
b\approx \frac{\ln(g)}{g}
\end{align}
\subsection{Small SNR Limit}
\label{smgsub}
Now we consider the case where the SNR is small, i.e., $g\to 0^+$. 

In order to solve for the optimal boundary, we treat $g$ as a perturbation, and observe the Taylor series of the quantization boundary $b$:
\begin{align}
b=b_0+b_1 g^2+b_2 g^4...
\end{align}

Then we can perform Taylor expansion on (\ref{2-3}) to obtain these coefficients. 

We define soft bit $\alpha$ for convenience:
\begin{align}\label{sftb}
\alpha=\frac{Q(b-g)-Q(b+g)}{Q(b-g)+Q(b+g)}
\end{align}

We simplify (\ref{2-3}) to reduce the complexity of the expansion:
\begin{align}
\label{splf}
\cosh(bg) \ln(1-\alpha^2)+\sinh(bg) \ln(\frac{1+\alpha}{1-\alpha})=0
\end{align}

The first-order expansion of (\ref{sftb})(\ref{splf}) indicates\footnote[1]{$Q'(x)$ represents $\frac{dQ(x)}{dx}$.}:
\begin{equation}
\label{1st}
\left\{\begin{array}{@{}ccc@{}}
-\alpha^2+ 2\alpha b_0 g =0\\
\\
\alpha=-g\frac{Q'(b_0)}{Q(b_0)}
\end{array}
\right.
\end{equation}

Equations (\ref{1st}) can be solved numerically, and the solution is $b_0\approx 0.6120$.

We also derive the second-order term, in appendix \ref{ap2}, which can be expressed as:
\begin{align}
b_1&=-\frac{1}{6}b_0
\end{align}

Theoretically, we can continue this process and solve for the solution up to an arbitrary order of significance; however, the complexity of the calculation increases dramatically, and the solutions become less valuable due to overfitting, which will be discussed later.
\label{large}

\subsection{Comparison with Numerical Results}

So far we have derived the asymptotic solutions of the optimal quantization threshold, summarized below:

Large SNR case: $b_L=\frac{\ln(g)}{g}$

Small SNR case (up to 1st order): $b_S\approx 0.6120$

Small SNR case (up to 2nd order): $b_{S2}\approx 0.6120 (1-\frac{g^2}{6})$

We extend these solutions to the entire range of the SNR as the approximated schemes. To eliminate the $b<0$ case, we replace negative boundary values with $0$.

Now we compare these approximations with the numerical solution.

\begin{figure}[ht!]
	\centering
	\includegraphics[width=120mm]{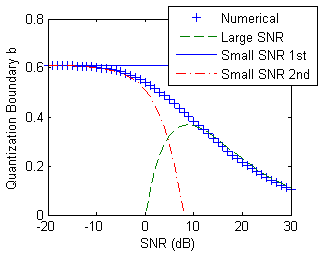}
	\caption{The quantization threshold against SNR.}
	\label{arm:fig3}
\end{figure}

\begin{figure}[ht!]
	\centering
	\includegraphics[width=120mm]{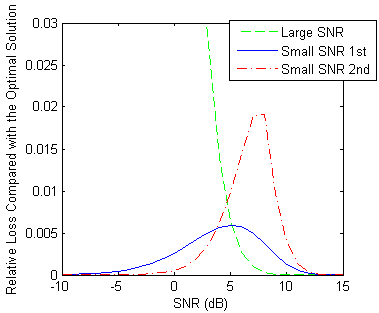}
	\caption{The relative loss of capacity against SNR. }
	\label{arm:fig4}
\end{figure}

We plot the quantization thresholds with respect to the SNR in Fig. \ref{arm:fig3}. We also calculate the relative difference of the capacity compared with the numerical solution and plot the curves in Fig. \ref{arm:fig4}.  

We can see that all three approximations converge to the numerical result in their corresponding region, for both quantization threshold and channel capacity, as we expected. 

More importantly, the 1st-order small SNR approximation provides minimal capacity loss on the entire range of SNR. The intuition is that the mutual information at large SNR is not sensitive to the quantization threshold; so, as long as a scheme satisfies the small SNR approximation, and does not overfit in the large SNR region, we can expect the loss to be reasonably small.

Thus we can also expect that the 1st-order small SNR approximation provides minimal loss for arbitrary quantization constraints when using 2-PAM modulation.

For simplicity, in the rest of the paper we call this approximation the \textit{small SNR approximation}.

\section{A Generalized Approach to Calculate the Small SNR Approximation}
\label{gen}

In this section, we generalize the small SNR approximation to the case that the number of quantization intervals can be arbitrarily large.

Instead of solving the optimality equations, we can first derive the 1st-order approximation of capacity in the small SNR case, and then generate an effective cost function based on the approximation.

The small SNR limit of the channel capacity has been discussed in \cite{ibragimov1972weak,prelov1993asymptotic} for both scalar channels and vector channels. For a scalar channel with transition density function $f(y|x)$, the approximation can be conveniently represented using the Fisher information:
\begin{align}
\label{smallcap}
I(X;Y)\approx \frac{Var(X)}{2}\mathcal{I}(0)
\end{align}

where $\mathcal{I}(x)$ is defined as:
\begin{align}
\mathcal{I}(x)=\int\frac{1}{f(y|x)}(\frac{\partial f(y|x)}{\partial x})^2 dy \text{.}
\end{align}

For simplicity, we make the following substitution:
\begin{align}
L(y|x)=\frac{\partial \ln f(y|x)}{\partial x} \text{.}
\end{align}

Then the Fisher information can be conveniently represented as:
\begin{align}
\label{noq}
\mathcal{I}(x)=\mathrm{E}_{Y|X}(L(y|x)^2) \text{.}
\end{align}

We can easily generalize the equation for the quantized channel:
\begin{align}
\label{qi}
\mathcal{I}_{quant}(x)=\mathrm{E}_{Z|X}(\mathrm{E}_{Y|Z,X}(L(y|x))^2) \text{.}
\end{align}

Then we consider using the relative loss of the capacity $R$, defined below, as the cost function.
\begin{align}
\label{rel}
R=1-\frac{I(X;Z)}{I(X;Y)}
\end{align}

With (\ref{smallcap})(\ref{noq})(\ref{qi}), the  relative loss of the capacity in the small SNR can be represented as:
\begin{align}
R&=1-\frac{\mathcal{I}_{quant}(0)}{\mathcal{I}(0)}\\
&=\frac{\mathrm{E}_{Z|X}(\mathrm{Var}_{Y|Z,X}(L(y|0)))}{\mathrm{E}_{Y|X}(L(y|0)^2)}
\end{align}

Now we look at the special case of Gaussian channels. We can easily derive that $L(y|0)=gy$. Then our cost function becomes:
\begin{align}
R&=\mathrm{E}_{Z}(\mathrm{Var}_{Y|Z}(y))
\end{align}

Hence the problem is completely reduced to an ordinary MSE quantization problem on a unit Gaussian variable.

A list of solutions, represented using the quantization thresholds, are included in Table \ref{tb1}, The results agree with those given in \cite{max1960quantizing}.

\begin{table}[ht!]
	\footnotesize
	\renewcommand{\arraystretch}{1.3}
	\caption{The small SNR optimal quantization thresholds for matched decoding }
	\centering
	
	\begin{tabular}{|c|l|}
		\hline
		Quantization Regions &  Locations of Positive Quantized Boundaries   \\\hline
		3& 0.6120 \\\hline
		4&  0.9816\\\hline
		5&  0.3823, 1.2443\\\hline
		6&  0.6589, 1.4468\\\hline
		7&  0.2803, 0.8744, 1.6107\\\hline
		8&  0.5005, 1.0499, 1.7479\\\hline
		9&  0.2218, 0.6812, 1.1976, 1.8655 \\\hline
		10&  0.4047, 0.8338, 1.3246, 1.9682\\\hline
		11& 0.1837, 0.5599, 0.9656, 1.4357, 2.0592\\\hline
		12& 0.3401, 0.6943, 1.0812, 1.5344, 2.1407\\\hline
	\end{tabular}
	\label{tb1}
\end{table}

Note that in the general case when the signals are not normalized, additional gain should be applied to these results. So the actual thresholds based on the small SNR approximation are proportional to the standard deviation of noise.

Now we compare the performance of our proposed solution with those of the numerical solution and the two traditional MSE quantization schemes introduced in \cite{shahid2008distributed}.

\begin{figure}[ht!]
	\centering
	\includegraphics[width=120mm]{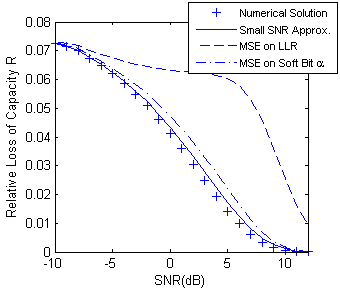}
	\caption{Comparison between different sub-optimal schemes.}
	\label{plt24}
\end{figure}

We plot the relative loss of capacity with respect to the SNR in Fig. \ref{plt24}, in the case where the channel output is quantized into $5$ possible values.  We observe that the small SNR approximation has strictly better performance than the MSE quantizations.


\begin{figure}[ht!]
	\centering
	\includegraphics[width=120mm]{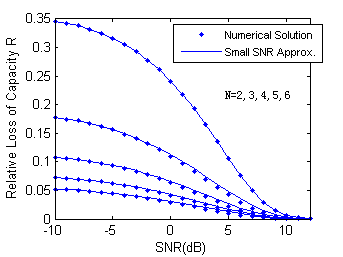}
	\caption{Performance of small SNR approximation under different quantization constraints.}
	\label{xx}
\end{figure}

We also make a plot that compares the relative capacity loss of the small SNR approximation and the optimal solution with different numbers of quantization regions (Fig. \ref{xx}). The plot shows that our approximation has a stable performance under different quantization constraints.  

\section{Asymptotic Loss of Capacity in High Rate Limit }
\label{largen}

In this section we study the trade-off relationship between the capacity loss and the resolution constraint in the high rate limit, i.e., $N\to +\infty$.

We make the following assumptions:
1. Each quantized output maps to a single interval.
2. $N$ is large enough so that, within a single interval, the conditional distribution $P_{X|Y}$ and its first order derivative $\frac{dP_{X|Y}}{dy}$ are nearly constant.

We first calculate the capacity loss due to quantization, denoted by $\Delta C$:
\begin{align}
\Delta C=\int \sum_x -f_Y P_{X|Y} ln\frac{P_{X|Z}}{P_{X|Y}}dy  \text{.}
\end{align}

Given that $P_{X|Y}$ is almost unchanged within each interval, we have $P_{X|Z}\approx P_{X|Y}$. Hence the following approximation holds:
\begin{align}
ln\frac{P_{X|Z}}{P_{X|Y}}\approx \frac{P_{X|Z}-P_{X|Y}}{ P_{X|Y}}-\frac{(P_{X|Z}-P_{X|Y})^2}{2 P_{X|Y}^2}
\end{align}

Then the capacity loss contributed from a single interval can be approximated as:
\begin{align}
\Delta C_z\approx \sum_x P_Z \frac{\mathrm{Var} (P_{X|Y};Z)}{2 P_{X|Z}}
\end{align}

Again, given the linear approximation of $P_{X|Y}$ for each single interval, the capacity loss can be represented as a weighted distortion function about $Y$:

\begin{align}
\Delta C&=\int  \sum_x \frac{f_Y}{2P_{X|Y}}(\frac{dP_{X|Y}}{dy})^2 \mathrm{Var} (Y;Z) dy
\end{align}

We slightly generalize the minimum distortion formula in the unweighted case, which was stated as equation (13) in \cite{panter1951quantization}, and obtain the minimum weighted distortion:

\begin{align}
\Delta C_{min}=\frac{1}{12N^2} (\int \sqrt[3]{\sum_x \frac{f_Y}{2 P_{X|Y}}(\frac{dP_{X|Y}}{dy})^2}dy)^3  \text{.}
\end{align}

The above equation can be simplified under the case of normalized Gaussian channels:

\begin{align}
\Delta C_{min}= \frac{g^2}{24N^2}(\int \sqrt[3]{{f_Y}\mathrm{Var}(X;Y)}dy)^3
\end{align}

\begin{figure}[ht!]
	\centering
	\includegraphics[width=120mm]{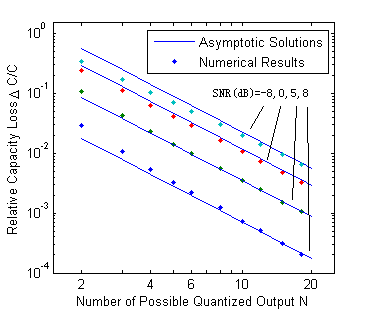}
	\caption{The relative capacity loss against $N$.}
	\label{fg7}
\end{figure}

Now we can compare the asymptotic solution with the numerical result in the case of 2-PAM modulation. We plot the relative loss $R$, defined in (\ref{rel}), for both results, in Fig. \ref{fg7}.

The numerical solutions quickly converge to the asymptotic curves; thus the ``inverse-square law'' can be used as a low-cost estimation of the capacity loss.

Finally, the approximated relative loss in the small SNR case can be easily calculated and expressed: 
\begin{align}
\lim_{g\to 0}R  &\approx \frac{1}{N^2}\lim_{g\to 0} \frac{\frac{g^2}{24}(\int \sqrt[3]{f_Y}dy)^3}{\frac{g^2}{2}}\\
&\approx \frac{\sqrt{3}\pi}{2}\frac{1}{N^2}
\end{align}

We conjecture that the relative loss of capacity monotonically decreases with respect to SNR, based on the observation from Fig. \ref{fg7}. If the conjecture is true, then the above formula gives an extremely simple estimator of the relative capacity loss in the worst case condition. 

\section{Summary}
\label{concl}

We examined the optimal quantization problem for a scalar Gaussian channel. Three main results are derived in this chapter:

\begin{enumerate}
	
	\item We proved that the optimal quantization regions for K-PAM Gaussian channels consist of a finite union of intervals. Especially in the 2-PAM case, the region that corresponds to any quantized output is proved to be a single interval. This result corresponds to the single-interval assumption in the MSE quantization case, and simplifies the discussion of the capacity maximization quantization problem.
	
	\item We showed that a linear approximation of the optimal quantization thresholds can be derived based on the asymptotic solutions in the small SNR case. The scheme gives no more that 0.6\% of relative capacity loss for 2-PAM modulated Gaussian channels. This quantization scheme can be implemented with low complexity (e.g. a linear analog gain control) when the receiver measures the channel condition and adapts its quantization strategy correspondingly.
	
	\item We derived a simple estimator of the relative capacity loss, based on the asymptotic solution in the high rate limit. For 2-PAM modulation, the worst case loss is $\sqrt{3}\pi/2N^2$, if the conjecture that the relative loss of the capacity is maximized at small SNR is true. This estimator gives a straightforward trade-off relationship between the system performance and the quantizer resolution. It also simplifies the system design by simplifying the process of picking the proper quantizer resolution.  
	
\end{enumerate}




\chapter{Optimal Quantizer for Mismatched Decoding with 2-PAM Modulation}
\label{cp:mm}
This chapter generalizes the asymptotic approximations we derived in chapter \ref{cp:matched} to the case of mismatched decoding. We derive the asymptotic solutions of the quantization thresholds, and study the trade-off between the loss of the mismatched capacity and the number of bits representing the decoding metrics.     

\section{Problem Overview}

We consider a discrete memoryless channel (DMC), with input alphabet $\mathcal{X}$, output alphabet $\mathcal{Y}$, and transition probabilities $P_{Y|X}$. 

Given a transmission of a coded message with length $n$, the receiver aims to reconstruct the transmitted string $X^n$, which is known to be selected from a given codebook $\mathcal{M}$, based on the channel output $Y^n$.

We suppose the receiver is equipped with a set of symbol-to-symbol decoding metrics $q(x,y)$, such that the receiver makes the decision by picking the string that maximizes the sum of the metrics. 

\begin{align}
\hat{X}^n(Y^n)=\mathop{argmax}_{X^n\in\mathcal{M}} \{\sum_{i=1}^{n}{q(x_i,y_i)}\}  
\end{align}

The maximum likelihood decision can be implemented by making the decoding metrics satisfy the following condition. 


\begin{align}
q(x,y)=A  \ln f_{Y|X}(y|x)+B(y)
\end{align}

\noindent where $A$ is a positive constant and $B(y)$ is an arbitrary function. 

However, in many cases, the decoding metrics in a practical system can only take integers from a finite set. This prevents the optimal solution from being implemented. The corresponding maximum achievable rate is referred to as the mismatched capacity $C'$.

In this chapter, we restrict our discussion to 2-PAM Gaussian channels and study the quantization problem under this scenario. 

We denote the quantized output alphabet as $\mathcal{Z}$, each corresponding to a possible pair of decoding metrics $q(x,z)$. For convenience, we define the relative metric $q(z)$ as follows.

\begin{align}
q(z)=q(1,z)-q(-1,z)
\end{align}

We restrict our quantizer to be symmetric, that is

\begin{align}
q(Z(y))=-q(Z(-y))
\end{align}
We want to pick the quantization function $Z(y):\mathcal{Y}\to \mathcal{Z}$ that maximizes the mismatched capacity.

\section{Background and Numerical Optimization}
\label{mbg}
The mismatched capacity $C'$ for a binary input discrete output channel $P_{Z|X}$ can be represented using the generalized mutual information $I_{GMI}(Z;X)$ \cite{balakirsky1995converse}.
\begin{align}
\label{mcap}
C'=\mathop{\max}_{P_X} I_{GMI}(X;Z)
\end{align}
\noindent where the generalized mutual information $I_{GMI}(X;Z)$ is defined as:
\begin{align}
I_{GMI}(X;Z)=\mathop{\min}_{P'_{X|Z}} \{H(X)-H_{P'}({X|Z})\}
\end{align}
\begin{align}
\text{s.t.} && \sum_z{P_Z{(z)} P'_{X,Z}(x,z)}&=P_X(x)&\\
&&\sum_{x,z}{P_Z{(z)} P'_{X,Z}(x,z) q(x,z)}&\geq \mathrm{E}[q(x,z)]&
\end{align}
Because it is non-trivial to maximize (\ref{mcap}) directly, an equivalent expression is commonly used for optimization problems.
\begin{align}
\label{mcap2}
C'=\mathop{\max}_{P_X,f_X,\alpha\geq0} H(X)+\sum_{x,z}P_{X,Z}(x,z)\ln(\frac{f_x e^{\alpha q(x,z)}}{\sum_{x'}f_{x'} e^{\alpha q(x',z)}})
\end{align}
For binary input symmetric channels, the capacity is optimized when $P_X$ is uniform and $f_X$ is a constant vector \cite{binshtok1999integer}.
\begin{align}
\label{mcaps}
C'_{Sym}=\log 2+\mathop{\max}_{\alpha\geq0}\sum_{x,z}P_{X,Z}(x,z)\ln(\frac{ e^{\alpha q(x,z)}}{\sum_{x'} e^{\alpha q(x',z)}})
\end{align} 
Then, for the optimized quantization scheme, $\exists \alpha\geq0$ such that the assignment of the decoding metrics satisfies
\begin{align}
\label{mnes}
Z(y)=\mathop{argmax}_z{P_{Y|X}(y|1) \ln(\frac{1}{1+e^{-\alpha q(z)}})+P_{Y|X}(y|-1) \ln(\frac{1}{1+e^{\alpha q(z)}})}
\end{align} 
From (\ref{mnes}), given any two quantized outputs $z_1$,$z_2$, the boundary of their quantization regions (if it exists), denoted by $b$, satisfies
\begin{align}
\label{bd}
\frac{P_{Y|X}(b|1)}{P_{Y|X}(b|-1)}=-\ln(\frac{1+e^{\alpha q(z_1)}}{1+e^{\alpha q(z_2)}})/\ln(\frac{1+e^{-\alpha q(z_1)}}{1+e^{-\alpha q(z_2)}})
\end{align} 
So far, we have derived that the optimal quantization regions only depends on the scalar parameter $\alpha$. The derivative of the maximum mismatched capacity as a function of $\alpha$ can be easily calculated. 

\begin{align}
\label{pa}
\frac{dC'(\alpha)}{d\alpha}=\frac{\partial C'(\alpha,Z(\cdot))}{\partial\alpha}=\sum_{z}{\frac{P_{Z|X}(z|1)q(z)}{1+e^{\alpha q(z)}}}
\end{align}
Then the numerical solution of the optimal quantization scheme can be calculated by any $1$D peak-finding algorithm.

\section{Optimal Quantization for 2-PAM Gaussian Channels}

We start by formally labeling the quantized outputs based on their relative metrics, and also labeling their quantization thresholds.

Practically, a quantized output with relative metric of $0$ gives no extra cost in the size of the decoder implementation, because no action is required in the decoding process. Hence, we assume there always exists a zero metric output in our discussion, and we label this quantized output as $0$.

For all the quantized outputs with a positive relative metric, we sort their metrics in an increasing order and label them from $1$ to $K$ ($K$ represents the number of such quantized outputs). Based on the symmetric quantizer assumption, we label their negative counterparts from $-1$ to $-K$. 

When a Gaussian channel is used, it is easy to verify that the relative decoding metric $q(Z(y))$ is an increasing function of the channel output $y$ based on (\ref{bd}), given that the quantization is optimal. 

\begin{figure}
 	\centering
 	\includegraphics[width=150mm]{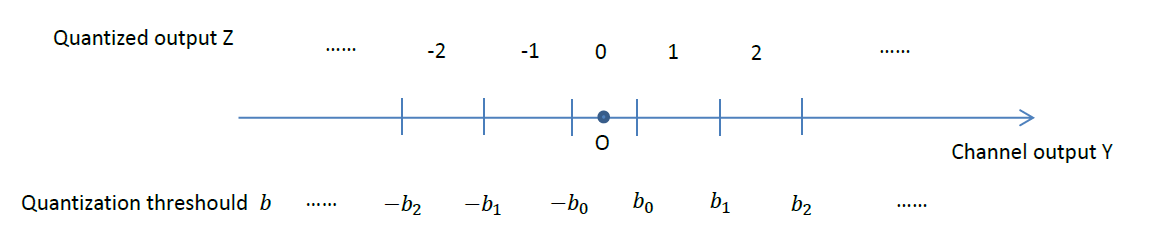}
 	\caption{An example of a mismatched quantization.}
 	\label{arm:fig31}
 \end{figure}

So the optimal quantization matches the quantized outputs to single intervals, and we can conveniently label the quantization thresholds using $\pm b_0,\pm b_1,...,\pm b_{K-1}$ (see Fig. \ref{arm:fig31}).   

The quantization intervals in this case can be expressed as
\begin{align}
\label{gbd}
b_i=\frac{1}{2g}\ln(-\ln(\frac{1+e^{\alpha q(i)}}{1+e^{\alpha q(i+1)}})/\ln(\frac{1+e^{-\alpha q(i)}}{1+e^{-\alpha q(i+1)}}))
\end{align} 

\noindent where $\alpha$ is the quantization parameter introduced in section \ref{mbg}.

\section{Asymptotic solutions}

In this section, we solve for the optimal quantization thresholds when the SNR is small or large for arbitrary decoding metrics. The solutions are fully determined by (\ref{pa}) and (\ref{gbd}).

\subsection{Large SNR Case}

In the large SNR case, $g\to+\infty$. We expect the optimal  quantization parameter $\alpha$ to grow to $+\infty$ so that the thresholds derived from (\ref{gbd}) provide non-zero solutions, which give a better capacity compared with a hard quantizer.

In this case, the thresholds can be approximated as
\begin{align}
\label{lp}
b_i=\frac{\alpha q_i}{2g}+o(1)
\end{align} 
 
Then, by applying $\frac{dC'(\alpha)}{d\alpha}=0$ to (\ref{pa}), we can derive the optimal quantization scheme for large SNR (for brevity, we put the derivation in Appendix \ref{ap3}).

\begin{align}
\left\{\begin{array}{@{}ccc@{}}
\ \alpha&=\frac{g^2}{2q_K}+O(\ln g)\\
\\
\ b_i&=\frac{gq_i}{4q_K}+o(1)
\end{array}
\right.
\end{align}

\subsection{Small SNR Case}

In the small SNR case, $g\to 0$. We expect that the optimal quantization parameter $\alpha$ decays to $0$. 

The optimal quantization thresholds can be approximated as
\begin{align}
\label{mbd}
b_i=\frac{\alpha (q(i)+q(i+1))}{4g}
\end{align}

We can also expand (\ref{pa}) about $g$ and $\alpha$.
\begin{align}
\label{smp}
\frac{dC'(\alpha)}{d\alpha}=\mathrm{E}_{Y}(\frac{L(y|0)q(z)}{2}-\frac{\alpha q(z)^2}{4})
\end{align}

\noindent where function $L(y|x)$ is defined as
\begin{align}
L(y|x)=\frac{\partial \ln f(y|x)}{\partial x}
\end{align}

When a Gaussian channel is used, $L(y|0)=gy$, the optimal quantization parameter satisfies the following equation.
\begin{align}
\label{malp}
\frac{\alpha}{g}=\frac{E_Y(2q(z)y)}{E_Y(q(z)^2)}
\end{align}

Essentially, the solution of the optimal quantization thresholds defined by (\ref{mbd}) and (\ref{malp}) is independent of the channel gain $g$. Thus we can calculate the thresholds numerically. 

Interestingly, the optimal quantization is uniform, when integer decoding metrics (i.e., $q(i)=i$) are used. Later in this chapter, we will see that the small SNR approximation gives negligible loss on the entire range of SNR, similar to the matched decoding quantization case. This explains the fact that uniform quantizers, which are implemented on most digital communication devices, do not give significant performance loss compared to non-uniform quantizers, even if the optimal solution is in general non-uniform.

In Table \ref{tb2}, we list the optimal quantization thresholds for integer metric quantizers with different quantization region numbers (defined as $N=2K+1$). For brevity, we include only the largest quantization threshold $b_{K-1}$.

\begin{table}[ht!]
	\footnotesize
	\renewcommand{\arraystretch}{1.3}
	\caption{The small SNR optimal quantization thresholds for mismatched decoding }
	\centering
	
	\begin{tabular}{|l|c|c|c|c|c|c|}
		\hline
		Quantization Region Number &  3  & 5& 7& 9& 11& 13  \\\hline
		Largest Quantization Threshold& 0.6120 & 1.2645& 1.6269 & 1.8683&  2.0460&  2.1846 \\\hhline{|=|=|=|=|=|=|=|}
		Quantization Regions & 15 &  17  & 19& 21& 23& 25  \\\hline
		Largest Quantization Threshold&   2.2975 & 2.3922& 2.4735 & 2.5445&  2.6074 & 2.6638\\\hhline{|=|=|=|=|=|=|=|}
		Quantization Regions & 27 &  29  & 31& 33& 35& 37  \\\hline
		Largest Quantization Threshold&   2.7148 & 2.7614& 2.8042 & 2.8437&  2.8805 & 2.9148\\\hline
	\end{tabular}
	\label{tb2}
\end{table} 

Remarkably, the optimal quantization thresholds for a 3-region quantizer have the same values as the solution we derived in the matched decoding case. This outcome is expected because at 3-region quantization, any symmetric decoding metrics are equivalent to matched metrics.

\section{Comparing with Numerical Solutions}

We compare the approximation schemes we derived from the asymptotic solutions to the numerical result, in the case of a 5-region quantizer with integer decoding metrics.

\begin{figure}[ht!]
	\centering
	\includegraphics[width=120mm]{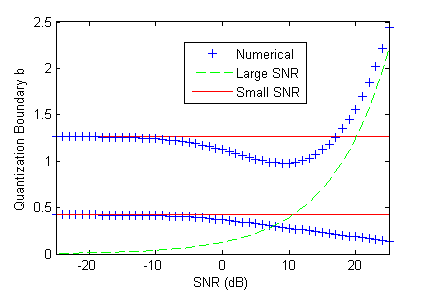}
	\caption{The mismatched quantization thresholds against SNR.}
	\label{mmb_ne5}
\end{figure}

\begin{figure}[ht!]
	\centering
	\includegraphics[width=120mm]{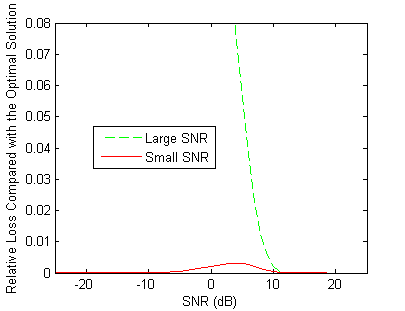}
	\caption{The relative loss of mismatched capacity against SNR.}
	\label{mmc_ne5}
\end{figure}

In Fig. \ref{mmb_ne5}, we plot the quantization thresholds against the SNR, and in Fig. \ref{mmc_ne5}, the relative loss of the mismatched capacity due to the asymptotic approximation.

Similarly to the matched quantization case, we observe that all the approximations converge to the numerical result in their corresponding region. In addition, the capacity loss produced from the small SNR approximation is extremely small, so that this approximated scheme is almost strictly better than the large SNR approximation, even in the large SNR case.  

\begin{figure}[ht!]
    	\centering
    	\includegraphics[width=120mm]{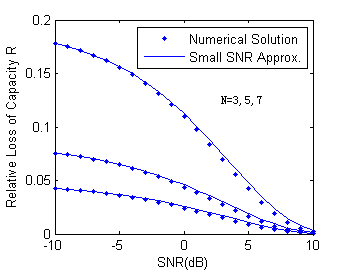}
    	\caption{Performance of small SNR approximation under different quantization constraints.}
    	\label{gmi_up}
    \end{figure}

We also compare the relative capacity loss from the small SNR approximation and the optimal solution with different quantization constraints. In Fig. \ref{gmi_up}, we plot the capacity loss due to quantization for integer metric decoding when the number of quantization regions equals 3, 5, or 7. We can see that the small SNR approximation has stable performance.
    
\section{Trade-off between Capacity Loss and Quantization Resolution}

Finding the trade-off relationship between the capacity loss and the quantization resolution in the mismatched case is also a practically useful topic. Detailed analysis shows that the relative loss of mismatched capacity for integer metric decoding grows at the speed of $\frac{\ln N}{N^2}$ (see Appendix \ref{ap4}). However, this high-rate approximation is not a good estimator of the capacity loss, since the convergence to this asymptotic solution requires $\sqrt{\ln N}\gg1$, which in practice is not true.

\begin{figure}[ht!]
	\centering
	\includegraphics[width=120mm]{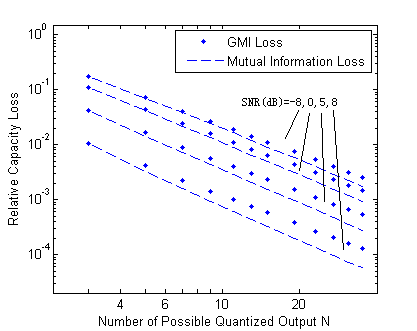}
	\caption{The trade-off between capacity loss and quantization resolution for matched and mismatched decoding.}
	\label{tradeoff_gmi}
\end{figure}

We also plot the trade-off curves for the matched and mismatched decoding case based on the numerical solutions (see Fig. \ref{tradeoff_gmi}). We can observe that the additional capacity loss due to the mismatched decoding is increasingly higher, which can be explained by the additional $\ln N$ factor in the high rate approximation formula.

\section{Summary}

We studied the optimal quantization problem for mismatched decoding in this chapter. 

We showed that the optimal quantization thresholds using small SNR approximation are still linear functions of the standard deviation of the noise, and they still give negligible loss on the mismatched capacity. 

We also showed that the relative loss of the mismatched capacity decays more slowly when we increase the resolution of the quantizer, compared with the matched decoding case. However, the asymptotic solution of the capacity loss no longer well approximates the numerical result; thus, finding a simple form estimator for the capacity loss is more difficult. 
\chapter{Mismatched Decoding for 8PSK Modulation}
\label{cp:bt}

In this chapter, we explore the mismatched decoding problem in the case of 8PSK bit interleaved modulation.

We first calculate the optimal decoding metrics assignment that maximizes the generalized mutual information or the general cutoff rate, and compare their error-rate performances. Then we introduce a low-complexity LLR decomposition algorithm for the 8PSK modulation, and show that the proposed demodulation scheme implemented on a commercial chip improves its performance. 


\section{Problem Definition}

The bit interleaved coded modulation, invented by Zehavi \cite{zehavi19928}, describes a scenario where a pair of bit interleavers are added to the signal chain of communication devices, so that each received symbol has to be decomposed to several bit metrics, and the decoder makes the decision correspondingly. 

Assigning the bit metrics to minimize the probability of getting an incorrect message for modulation types like 8PSK or 4PAM is non-trivial, because the a posteriori probability distributions of the bits are not independent given the received symbols.

Alternatively, a common approach to solving this problem is to consider it as a mismatched decoding scheme, and to maximize objective functions like the GMI or the GCR \cite{jalden2010generalized,martinez2009bit,nguyen2011bit}.

\begin{figure}[ht!]
	\centering
	\includegraphics[width=80mm]{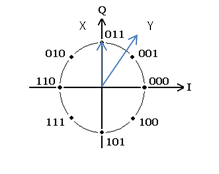}
	\caption{8PSK bit-constellation assignment}
	\label{8psk}
\end{figure}

Here, we focus on the case of 8PSK Gaussian channel, with its bit-constellation assignment specified in Fig. \ref{8psk}.

We assume that each transmitted symbol $x$ is converted from three input bits $b_1$, $b_2$, and $b_3$, and each received symbol $y$ is decomposed into three bit metrics $q_1$, $q_2$, and $q_3$. Based on the definition of BICM, the metrics $q(x,y)$ that are used by the decoder can be expressed using the three bit decoding metrics.

\begin{align}
\label{dec}
q(x,y)&=\sum_{i=1}^3 {q_i(b_i,y)}\\
	&=\sum_{i=1}^3 {q_i(y) \ \mathbf{1}(b_i=0)}
\end{align}

To simplify our discussion, we ignore the integer metric constraint and assume the bit metrics can be arbitrary real numbers.

\section{Decoding Metrics that Maximize GCR or GMI}

Now we consider the GMI and the GCR of BICM channels under the uniform input distribution constraint, and we compare the error rate performance of the optimal bit metrics assignments that maximizes these objective functions.  
\subsection{Decoding Metrics that Maximize GMI}

First we derive the optimal bit metric assignment that maximizes the generalized mutual information. This solution has also been discussed in \cite{martinez2009bit}. 

From section \ref{mbg}, we recall that the generalized mutual information can be expressed as
\begin{align}
I_{GMI}(X,Y)=\mathop{\max}_{f_X,\alpha\geq0} H(X)+\sum_{x,y}P_{X,Y}(x,y)\ln(\frac{f_x e^{\alpha q(x,y)}}{\sum_{x'}f_{x'} e^{\alpha q(x',y)}})
\end{align}

For an 8PSK modulated channel with a symmetric input distribution, the above formula can be further simplified into
\begin{align}
I_{GMI}(X,Y)=3\log 2+\mathop{\max}_{\alpha\geq0}\sum_{x,y}P_{X,Y}(x,y)\ln(\frac{ e^{\alpha q(x,y)}}{\sum_{x'} e^{\alpha q(x',y)}})
\end{align} 

When bit-interleaved coded modulation is also used, by applying (\ref{dec}), the generalized mutual information equals the sum of the GMI of three independent binary input channels.
\begin{align}
I_{GMI}(X,Y)&=\mathop{\max}_{\alpha\geq0}\sum_{i=1}^3 (\log 2+\sum_{b_i,y}P_{B_i,Y}(b_i,y)\ln(\frac{ e^{\alpha q(b_i,y)}}{\sum_{b_i'} e^{\alpha q(b_i',y)}}))
\end{align} 
Hence, the optimal bit metrics that maximize the GMI are proportional to the log-likelihood ratios of the marginal probability distribution of each bit.

\begin{align}
q_i(y)&=A \ln\frac{P_{B_i,Y}(0,y)}{P_{B_i,Y}(1,y)}\\
&=A \ln\frac{\sum_{x:b_i=0}P_{Y|X}(y|x)}{\sum_{x:b_i=1}P_{Y|X}(y|x)}
\end{align} 

\noindent where $A$ is an arbitrary constant.

For 8PSK Gaussian Channels, the log-likelihood ratio equals
\begin{align}
LLR_i&=\ln\frac{\sum_{x:b_i=0}e^{-\frac{||x-y||^2}{2}}}{\sum_{x:b_i=1}-\frac{||x-y||^2}{2}}\\
&=\ln\frac{\sum_{x:b_i=0}e^{x\cdot y}}{\sum_{x:b_i=1}e^{x \cdot y}}
\end{align}
  
\noindent where we use $||\cdot||$ to denote the norm of a vector, and use $\cdot$ to denote the dot product.

Practically, the max-log approximation $\ln\sum_{i}{x_i}\approx \mathop{\max}_i \ln(x_i)$ is frequently used in the calculation of the marginal LLR. 
 \begin{align}
 \label{appllr}
 LLR_i&\approx \mathop{\max}_{x:b_i=0}x\cdot y-\mathop{\max}_{x:b_i=1} x\cdot y
 \end{align}
 
 In this way, the approximated LLR can be calculated without an exponential function calculator.

\subsection{Decoding Metrics that Maximize GCR}

The general cut-off rate, derived as a cut-off rate in the mismatched decoding case, was defined in \cite{salz1995decoding}.

 \begin{align}
R_{GCR}=-\mathop{\max}_{\alpha>0}{\ln{\sum_{x}{P_{X,Y}(x,y)\frac{\sum_{x'} P_{X'}(x')e^{\alpha q(x',y)}}{e^{\alpha q(x,y)}}}}}
 \end{align}
 
 When $P_X$ is uniform, the formula can be simplified to 
 
 \begin{align}
 R_{GCR}=H(X)-\mathop{\max}_{\alpha>0}{\ln{\sum_{x,y}{P_{X,Y}(x,y)\frac{\sum_{x'} e^{\alpha q(x',y)}}{e^{\alpha q(x,y)}}}}}
 \end{align}
 
 Thus, finding the bit metrics that maximizes the GCR is equivalent to solving the equation below. 
  \begin{align}
  \label{tar}
 (q_1,q_2,q_3)=\mathop{argmin}_{q_1,q_2,q_3} \sum_{x}{P_{X,Y}(x,y)\frac{\sum_{x'} e^{\alpha q(x',y)}}{e^{\alpha q(x,y)}}}
  \end{align}
  
  For 8PSK modulated channels, although deriving the exact solution to (\ref{tar}) is non-trivial, a simple approximated solution exists, when the signal-to-noise ratio of the channel is large enough that the max-log approximation can be applied (see Appendix \ref{ape} for the derivation).
  \begin{align}
  & q_1=
  \left \{\begin{array}{@{}lll@{}}
  A\ LLR_1,&& LLR_1<LLR_2\\
  A\ (LLR_1-|\frac{LLR_3}{2}|\text{sgn}(LLR_1)),&& LLR_1>LLR_2
  \end{array}
  \right.\\
    & q_2=
    \left \{\begin{array}{@{}lll@{}}
    A\ LLR_2,&& LLR_1>LLR_2\\
    A\ (LLR_2-|\frac{LLR_3}{2}|\text{sgn}(LLR_2)),&& LLR_1<LLR_2
    \end{array}
    \right.\\
      & q_3=A\ LLR_3
      \end{align}
  \noindent where A is an arbitrary positive value.

  From (\ref{appllr}), we know this approximated solution can also be calculated without an exponential function calculator. Hence it can also be implemented on hardware with reasonable cost.
  
  \subsection{Error Rate Performances of the Two Demodulation Schemes}
  
  We compare the error rate performances of the two bit-metric assignments on a powerline communication system developed by Maxim Integrated. We simulate the operation of the chip in a model written in C, and we use time domain AWGN channels to evaluate its performance.  
  
\begin{figure}[ht!]
  	\centering
  	\includegraphics[width=140mm]{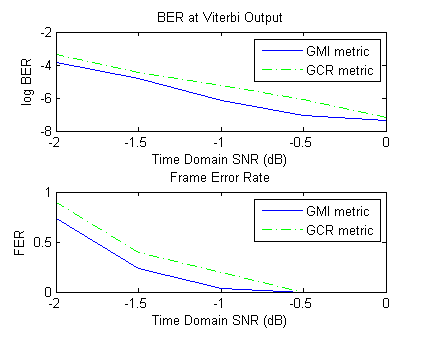}
  	\caption{The error rates for two different bit-metric assignments.}
  	\label{fer1}
  \end{figure}

  The system uses RSV code for error correction, so we plot the bit error rate at the Viterbi decoder's output, as well as the message error probability, against different time domain SNR (see Fig. \ref{fer1}).   
  
  From the plots, we can see that the bit-metric assignment that maximizes the GMI gives a better performance, whether the Reed-Solomon code is active or only the convolutional code is used. We also conclude that, although the general cut-off rate is a function that evaluates the performance of convolutional code based systems, choosing a demodulation scheme that maximizes the GCR does not necessarily minimize the error probability.







\section{Error Rate Performance of the GMI Maximization Demodulation Based on a Fast LLR Decomposition Algorithm}

The following text discusses the implementation of the GMI-based 8PSK demodulation algorithm, and the test result of its performance on a physical chip.

\subsection{Firmware Implementations of 8PSK Demodulation Algorithms} 
 The main task of calculating the optimal bit metrics that maximizes the GMI is to calculate the LLR for each bit. Several low-complexity approximations and their implementations have been discussed in \cite{barre2013polar,cheng2009structured,park2008low}.
 For 8PSK modulation, we propose a simple algorithm that calculates the max-log approximated LLRs as defined in (\ref{appllr}).
 
\subsubsection{Proposed Demodulation Algorithm}

From (\ref{appllr}), we know that in order to calculate the LLR for each bit, we need to first find the closest constellation points to the channel output $Y$ that correspond to bit values 0 and 1, and then calculate the difference of their dot products to $y$. 

\begin{figure}[ht!]
	\centering
	\includegraphics[width=80mm]{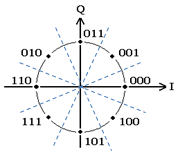}
	\caption{The 8-region devision of the channel output space for 8PSK demodulation.}
	\label{8pd2}
\end{figure}

We divide the complex plane into 8 regions (see Fig. \ref{8pd2}) so that, for each region, the selection of the closest constellation points remains the same. Thus the LLR calculation in each region can be directly implemented by calculating the dot product of Y and a fixed vector.

For each channel output $y$, we denote the closest constellation point that satisfies $b=0$ by $x_0$. Similarly we denote the counterpart by $x_1$. Then the approximated LLR equals
 \begin{align}
 LLR_i&\approx (x_0- x_1)\cdot y
 \end{align}
 
\begin{figure}[ht!]
	\centering
	\includegraphics[width=80mm]{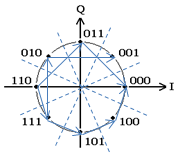}
	\caption{8 possible values of $(x_0- x_1)$}
	\label{8dot}
\end{figure}

 There are effectively 8 possible values of $(x_0- x_1)$ (see Fig. \ref{8dot}), thus the dot product can be implemented with no more than 4 fixed complex multiplications. 

At the same time, these dot products also provide information about which region contains the channel output $y$. Thus, no extra complex multiplication is needed to look up the regions. 

Now we propose a firmware algorithm that uses jump look-up to handle the LLR calculations in all 8 regions.

\begin{figure}[ht!]
	\centering
	\includegraphics[width=80mm]{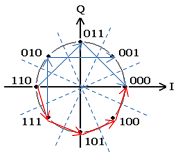}
	\caption{4 dot products (marked as red) required to be calculated in step 1.}
	\label{8m}
\end{figure}

\emph{Step 1:}
We calculate all 4 dot products marked in Fig. \ref{8m} using 2 complex multiplications. Because all those vectors are parallel to the boundaries of the $8$ regions, we can also identify the region that contains channel output from the multiplication results.

\emph{Step 2:}
For each region, in order to calculate the LLRs for all 3 bits, only one extra dot product operation is needed. Thus we use another complex multiplication to complete the calculation.     
\subsection{Error Rate Performance of the Proposed Algorithm} 

\begin{figure}[ht!]
	\centering
	\includegraphics[width=150mm]{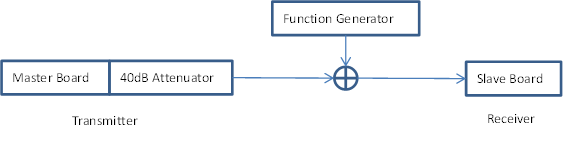}
	\caption{Board test configuration}
	\label{bt}
\end{figure}

We implemented the proposed algorithm on the powerline communication chip provided by Maxim Integrated, and tested the performance of different demodulation algorithms using the configuration described in Fig. \ref{bt}. A function generator is used to generate the noise, and we control the SNR by changing the noise amplitude.  

\begin{figure}[ht!]
	\centering
	\includegraphics[width=150mm]{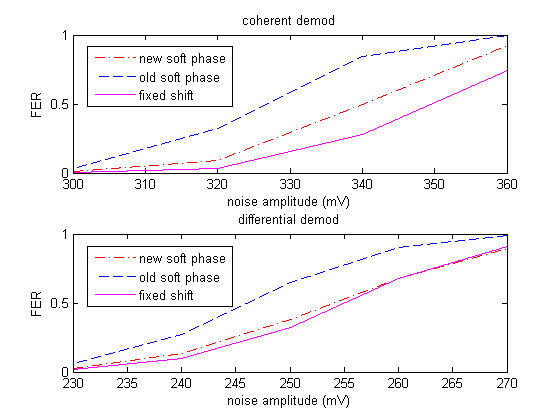}
	\caption{The error probability curves for three different demodulation algorithms}
	\label{fer2}
\end{figure}

We plot the error probability under different noise amplitudes, for three different demodulation algorithms (in Fig. \ref{fer2}). 

What we refer to as the old soft phase demodulation is the original demodulation scheme that was implemented on the powerline communication chip. It calculates the bit metrics based only on the phase of the received symbol. A direct replacement of the look-up table, which is referred to as the new soft phase demodulation, gives a better performance. The fixed shift demodulation is the scheme that we implemented in this thesis, which has the best performance among the three demodulation schemes according to the plot. Compared to the original demodulation scheme, the fixed shift gives a $0.6$dB improvement for coherent modulation, and a $0.3$dB improvement for differential modulation.   



\section{Summary}

In this chapter, we explored the problem of finding the optimal bit-metric assignment for 8PSK modulated channel.

First we showed that the marginal LLRs, are the bit metrics that maximize the generalized mutual information. This is also partially the reason that many BICM systems use it in the demodulator. However, this metric is not always optimal under different objective functions, e.g., the general cut-off rate. We derived the corresponding optimal bit metrics when max-log approximation holds, and they are simple linear functions that depend on the LLRs. 

We also developed an efficient algorithm that calculates the max-log approximated LLRs. The demodulation scheme based on this algorithm was implemented and tested on a physical chip. From the testing result, this scheme improves the system for $0.6$dB in the coherent mode and for $0.3$dB in the differential mode.

\chapter{Conclusion and Future Work}

\label{cp:cl}

Finding the optimal demodulation schemes is a practical problem attracting much research interest. Various objective functions have been proposed as the optimization metric. In this thesis, we explored the problem from multiple aspects, which include maximizing different objective functions under different scenarios, and deriving asymptotic solutions with low implementation cost that well approximate the numerical solutions. 

We started from the simplest case, where the modulation type of the signal is 2-PAM, and the channel output is quantized into finitely many regions. We developed approximate quantization schemes based on the small SNR asymptotic solutions of the optimal result. It gives negligible capacity loss in both the matched decoding and the mismatched decoding cases, while being simple to implement. We also studied the trade-off between the capacity loss and resolution of the quantizer. We showed that a very simple estimator of the relative loss of capacity can be derived in the matched decoding case using the high-rate limit approximation.

Then we considered a more practical case where not only is the channel output quantized, but the decoding metrics are also constrained to given values. We showed that the small SNR approximation of the quantization thresholds still works in this case, which provides an approximation scheme that can be directly used on most modern communication devices.   

Finally we looked into the demodulation problem for 8PSK modulated channels. We discussed the optimal bit-metric assignment for the GMI and the GCR. We also implemented a simple algorithm that decomposes the marginal LLRs based on the received symbol, which improves the performance of the chips where it was implemented.

In addition to the topics we have discussed, there are still many related problems that remain open.

\begin{enumerate}
		\item In the case of matched decoding quantization, we have shown that the approximate quantization scheme derived from asymptotic solutions gives negligible capacity loss on binary input channels. However, solving the asymptotic solutions in the general cases, e.g. when 8PSK modulation is used, is less trivial. Thus, extending this approximation approach to different modulation types remains to be researched.  
		
		\item In the analysis of the trade-off between the capacity loss and the quantizer resolution, we conjectured that the minimum relative loss of capacity is a decreasing function of the SNR for 2-PAM Gaussian channels. If this is true, we can derive a simple formula that estimates the worst case relative loss of the capacity in this scenario.  
		
		\item Although the mismatched capacity is equal to the GMI for binary input channels, it is not equal to the GMI lower bound, and its formula provided in \cite{somekh2014general} does not single-letterize in the general case. This makes it hard to maximize the actual mismatched capacity in many cases. Whether a simple formula exists for specific cases like 8PSK BICM channels, would be an interesting and valuable problem to research on.  
		

\end{enumerate}

\appendix
\chapter{Proof of Theorem \ref{tm1}}

\label{ap1}
\begin{proof}

First we apply the Gaussian channel condition to (\ref{req}):
\begin{align}
\label{lm0eq}
z(y)=\mathop{argmax}_{Z} \{\sum_X{P_X e^{\frac{-(gx-y)^2}{2}}ln(P_{X|Z})}\}
\end{align}

For $K$-PAM modulation, any possible channel input value $x$ can be represented as:
\begin{align}
x=x_i=x_0+id
\end{align}

Where $i$ takes value from $\{0,1,2,...,K-1\}$.

Then (\ref{lm0eq}) can be further simplified:
\begin{align}
z(y)=\mathop{argmax}_{Z} \{\sum_{i=0}^{K-1}{P_X e^{\frac{-(gx_i)^2}{2}+idgy}ln(P_{X|Z})}\}
\end{align}

We define variable $\phi=e^{d gy}$. Then the quantization intervals of each quantized output $z$ can be fully determined by a set of polynomial constraints:
\begin{align}
\label{polycons}
\sum_{i=0}^{K-1}{(a_{i,z}-a_{i,z'})\phi^i}\geq0
\end{align}
for every $z'\neq z$, with $a_{i,z}=P_Xe^{\frac{-(gx_i)^2}{2}}ln(P_{X|Z})$.

Since each polynomial constraint produces a set of disjoint intervals, and that the quantization intervals are the result of their intersection, the entire problem is reduced to a mathematical problem about counting the number of intervals.

So we first introduce some definitions in order to simplify the following discussion.

\begin{definition}
	The \textbf{boundaries} of a union of disjoint intervals $S=\cup_i [a_i,b_i]$ are defined as $\cup_i \{a_i,b_i\}$, denoted by $B(S)$.
\end{definition}

\begin{definition}
	The \textbf{inner boundaries} of a union of disjoint intervals $S$ and defined as $B(S)-\{S_{max},S_{min}\}$, denoted by $B'(S)$.
\end{definition}

For simplicity, we denote the number of elements in an finite set $B$ by $|B|$, and we denote the number of intervals in their union $S$ to be $|S|$

We then prove the following lemma:
\begin{lemma}
	\label{jlma}
	Given two arbitrary unions of disjoint intervals $S_1$,$S_2$. We have
	\begin{align}
	|B'(S_1\cap S_2)|\leq |B'(S_1)|+|B'( S_2)|
	\end{align}
\end{lemma}
\begin{proof}
	For an arbitrary set of disjoint intervals $S$, a necessary condition for a real number $x$ to belong to $B'(S)$ is that $x$ satisfies $S_{min}<x<S_{max}$. 
	
	Thus the following relation holds:
	\begin{align}
	B'(S_1\cap S_2) &=B(S_1\cap S_2)\nonumber\\
	&-\{S_{1,max},S_{1,min},S_{2,max},S_{2,min}\}
	\end{align}
	
	It is easy to prove that 
	\begin{align}
	B(S_1\cap S_2)\subseteq B(S_1)\cup B(S_2).
	\end{align}
	
	Then we have
	\begin{align}
	B'(S_1\cap S_2)&\subseteq (B(S_1)-\{S_{1,max},S_{1,min}\})\nonumber\\
	&\cup(B(S_2)-\{S_{2,max},S_{2,min}\})\\
	&=B'(S_1)\cup B'(S_2)
	\end{align}
	
	The following can thus be derived directly:
	\begin{align}
	|B'(S_1\cap S_2)|\leq |B'(S_1)|+|B'( S_2)|
	\end{align}
	
\end{proof}

Now we can look at the intervals number counting  problem we mentioned earlier.

From (\ref{polycons}), each single constraint generates a set of disjoint intervals. Since the constraints are polynomial on $\phi$ with degrees no larger than $K-1$, there are at most $K-1$ roots for each constraint. We know that the number of inner boundaries has to be even, so the upper bound on the number of inner boundaries from one constraint is $2\lfloor\frac{K-1}{2}\rfloor$. 

Since there are $N-1$ constraints for each quantized output $z$. We ultilize Lemma \ref{jlma} and conclude that $|B'(S(z))|\leq 2(N-1)\lfloor\frac{K-1}{2}\rfloor$.

It is straight forward to prove that $|B'(S)|=2(|S|-1)$. So the upper bound on the number of disjoint intervals that map to the same quantized output in the optimal quantization scheme can be represented as:
\begin{align}
|S(z)|\leq (N-1)\lfloor\frac{K-1}{2}\rfloor+1
\end{align}

\end{proof}
	
\clearpage
\newpage

\chapter{The Second Order Expansion for Small SNR Matched Quantization}
\label{ap2}

We solve for the second order term of the quantization threshold $b_1$, for the problem defined in \ref{smgsub}.

First we expand the soft bit $\alpha$ to the second order:
\begin{align}
\label{af2}
\alpha=-\frac{Q'(b_0)x+Q''(b_0)xb_1+\frac{1}{6}Q'''(b_0)x^3}{Q(b_0)+Q'(b_0)b_1+\frac{1}{2}Q''(b_0)x^2}
\end{align}

Based the properties of $b_0$ and the $Q$ function, the following equation holds:
\begin{align}
2b_0 Q(b_0)&=-Q'(b_0)\\
Q''(b_0)&=-b_0Q'(b_0)\\
Q'''(b_0)&=(b_0^2-1)Q'(b_0)
\end{align}

These equations can simplify the expression of $\alpha$:
\begin{align}
\label{simpa}
\alpha=-g\frac{1-b_0b_1 g^2+\frac{1}{6}(b_0^2-1)g^2}{-\frac{1}{2b_0}+b_1 g^2-\frac{1}{2}b_0g^2}
\end{align}

Now we expand (\ref{splf}) to the second order:

\begin{align}
\label{ss2}
&(1+\frac{(b_0 g)^2}{2})(-\alpha^2-\frac{\alpha^4}{2})\nonumber\\
&+(b_0+b_1g^2)g(1+\frac{(b_0g)^2}{6})(2\alpha+\frac{2\alpha^3}{3})=0
\end{align}

Plugging (\ref{simpa}) into (\ref{ss2}), we can solve for $b_1$:
\begin{align}
b_1&=-\frac{1}{6}b_0
\end{align}

\clearpage
\newpage

\chapter{Large SNR Approximation for 2PAM Mismatched Quantization}

\label{ap3}

by applying $\frac{dC'(\alpha)}{d\alpha}=0$ to (\ref{pa}), we have
\begin{align}
\label{mm:ls316}
\ln \sum_{z=-1}^{-K}{-\frac{P_{Z|X}(z|1)q(z)}{1+e^{\alpha q(z)}}}=\ln \sum_{z=1}^{K}{\frac{P_{Z|X}(z|1)q(z)}{1+e^{\alpha q(z)}}}
\end{align}

Using max-log approximation $\ln\sum_{i}{x_i}\approx \mathop{\max}_i \ln(x_i)$, \ref{mm:ls316} can be simplified to 

\begin{align}
\label{mm:ls317}
\ln {-P_{Z|X}(-1|1)q(-1)}=\mathop{\max}_{z>0}\{\ln P_{Z|X}(z|1)q(z)-\alpha q(z)\}
\end{align}

We can approximate the transition probability of the quantized channel using (\ref{logq}). 
\begin{align}
\label{pz}
\ln P_{Z|X}(z|1)\approx \mathop{\max}_{y:Z(y)=z} -\frac{(y-g)^2}{2}
\end{align}

Thus,
\begin{align}
\label{pzf}
\ln P_{Z|X}(-1|1)&\approx -\frac{g^2}{2}
\end{align}

For any $0<z<K$, the following inequality can be derived.
\begin{align}
\ln P_{Z|X}(z|1)\leq -\frac{(g-\min\{b_z,g\})^2}{2}
\end{align} 

Then using (\ref{lp}), we have
\begin{align}
\ln P_{Z|X}(z|1)q(z)-\alpha q(z)\leq -\frac{g^2}{2} +O(\ln g)
\end{align}

So \ref{mm:ls317} can be simplified to 

\begin{align}
-\frac{g^2}{2}&=\ln P_{Z|X}(K|1)-\alpha q(K)+O(\ln g)\\
&= -\frac{(g-\max\{b_{K-1},g\})^2}{2}-\alpha q(K)+O(\ln g)
\end{align}

The solution is 

\begin{align}
\left\{\begin{array}{@{}ccc@{}}
\ \alpha&=\frac{g^2}{2q_K}+O(\ln g)\\
\\
\ b_i&=\frac{gq_i}{4q_K}+o(1)
\end{array}
\right.
\end{align}

\chapter{High-rate Limit for Mismatched Quantization with Integer Metrics}

\label{ap4}

Given a 2-PAM Gaussian channel with a quantize function $Z(y)$, the loss of mismatched capacity equals

\begin{align}
\label{dc}
\Delta C=\mathop{\min}_{\alpha}\int f_y \sum_x P_{X|Y}(x|y) (\ln \frac{e^{xy}}{\sum_{x'} e^{x'y}}-\ln \frac{e^{\alpha q(x,z)}}{\sum_{x'} e^{\alpha q(x',z)}})   dy
\end{align}

When the number of quantization region is large, the optimal relative decoding metric $q(z)$ should match with the actual log-likelihood ratio, so that the loss of capacity can approach zero, i.e., 
\begin{align}
\alpha q(z)\approx 2gy
\end{align}
with probability of 1.
  
Thus, we can Taylor expand (\ref{dc}) and derive the capacity loss in the high-rate limit.
   
\begin{align}
\label{apmc2}
 \Delta C\approx \mathop{\min}_{\alpha} \int \frac{f_y (\alpha d-2gy)^2}{2(e^{gy}+e^{-gy})^2} dy
 \end{align} 
 
 The approximate loss of capacity, defined by (\ref{apmc2}), can be calculated by dividing the integral into 2 parts: $|2gy|<\alpha K$ and $|2gy|>\alpha K$. For convenience, we define this boundary to be $y_m=\frac{\alpha K}{2g}$.
 
The first part of the integral can be treated by assuming the length of the quantization intervals approaches to $0$ when the number of quantization regions $N$ is large. Thus the coefficients of the ``square error" function is almost constant within a interval.
\begin{align}
\label{int1}
\int_{|2gy|<\alpha K} \frac{f_y (\alpha d-2gy)^2}{2(e^{gy}+e^{-gy})^2} dy&\approx \int_{|2gy|<\alpha K} \frac{f_y \mathrm{Var}(2gy|Z=Z(y))}{2(e^{gy}+e^{-gy})^2} dy\\
&\approx \int_{-\infty}^{+\infty} \frac{f_y }{2(e^{gy}+e^{-gy})^2}\frac{(2gy_m)^2}{3N^2} dy
\end{align} 

To simplify the above equation, we substitute the coefficients that are independent of the quantizer parameters $\alpha$ and $N$. 
\begin{align}
\label{1sti}
A=\int_{-\infty}^{+\infty} \frac{2f_yg^2 }{3(e^{gy}+e^{-gy})^2}dy
\end{align} 

Then the first part of the integral can be simply expressed as $Ay_m^2/N^2$.

Before calculating the second part of the integral, we derive the following lemma.
\begin{lemma}
	\label{lm6}
	For any $x>0$, the following inequality hold.
	\begin {align}
	\frac{2-e^{-(x+\frac{1}{2})}(x^2+3x+\frac{13}{4})}{(x+\frac{1}{2})^3} e^{-\frac{x^2}{2}}\leq\int_0^{+\infty}e^{-\frac{(x+y)^2}{2}}y^2 dy\leq \frac{2}{x^3} e^{-\frac{x^2}{2}}&&
	\end {align}
\end{lemma}
\begin{proof}
	First we derive the upper bound
	\begin {align}
	\int_0^{+\infty}e^{-\frac{(x+y)^2}{2}}y^2 dy&\leq \int_0^{+\infty}e^{-\frac{x^2}{2}-xy}y^2 dy\\
	&=\frac{2}{x^3} e^{-\frac{x^2}{2}}
	\end {align}
	
	Then we derive the lower bound
	\begin {align}
	\int_0^{+\infty}e^{-\frac{(x+y)^2}{2}}y^2 dy&\geq \int_0^{1}e^{-\frac{x^2}{2}-(x+\frac{1}{2})y}y^2 dy\\
	&=\frac{2-e^{-(x+\frac{1}{2})}(x^2+3x+\frac{13}{4})}{(x+\frac{1}{2})^3} e^{-\frac{x^2}{2}}
	\end {align}
\end{proof}

From Lemma \ref{lm6} we obtained a convenient approximation
\begin{align}
\int_0^{+\infty}e^{-\frac{(x+y)^2}{2}}y^2 dy\approx \frac{2}{x^3} e^{-\frac{x^2}{2}} 
\end{align}
\noindent which simplifies the second part of the integration of the capacity loss.
\begin{align}
\int_{|2gy|>\alpha K} \frac{f_y (\alpha d-2gy)^2}{2(e^{gy}+e^{-gy})^2} dy&\approx \int_{y>y_m} \frac{4g^2}{\sqrt{2\pi}}  e^{-\frac{(g+y)^2}{2}} (y-y_m)^2 dy\\
&\approx \frac{8g^2}{\sqrt{2\pi}(g+y_m)^3}  e^{-\frac{(g+y_m)^2}{2}} 
\label{2ndi}
\end{align} 

Similarly, we make the following substitution for the coefficient of the second integral.
\begin{align}
B=\frac{8g^2}{\sqrt{2\pi}}  
\end{align}

Combining (\ref{1sti}) and (\ref{2ndi}), we obtain a simple formula that approximates the capacity loss.

\begin{align}
\Delta C\approx \mathop{\min}_{y_m} \{\frac{Ay_m^2}{N^2}+\frac{B}{(g+y_m)^3}  e^{-\frac{(g+y_m)^2}{2}} \}
\end{align} 

When $y_m\gg1$, with some simple analysis, the loss of capacity is minimized at

\begin{align}
y_m=\sqrt{4\ln N}+O(1)
\end{align} 

and the capacity loss equals 
\begin{align}
\label{finl}
\Delta C\approx \frac{4A\ln N}{N^2}+O(\frac{\sqrt{\ln N}}{N^2})
\end{align} 

So far, we have derived the minimum mismatched capacity loss for 2-PAM Gaussian Channels in the high-rate limit. In terms of using (\ref{finl}) as a estimator of the capacity loss, it requires $\sqrt{\ln N}\gg1$, which in practice is not true. So the approximate formula of capacity loss in the mismatched case can not be derived from this approach.
\chapter{Optimal Demodulation Maximizing GCR for 8PSK Gaussian Channel}

\label{ape}

Finding the optimal bit metrics that maximizes the GCR is essentially solving equaltion (\ref{tar}). Although finding an exact formula of the solution is generally difficult, we can instead derive an approximate result when max-log approximation can be used, i.e., solving the below equation instead.
  \begin{align}
  (q_1,q_2,q_3)=\mathop{argmin}_{q_1,q_2,q_3} \mathop{\max}_x{P_{X,Y}(x,y)\frac{\sum_{x'} e^{\alpha q(x',y)}}{e^{\alpha q(x,y)}}}
  \end{align}  

Specifically, for 8PSK Gaussian channel with the labeling of constellation points specified in Fig. \ref{8psk}, the above equation is equivalent to
        \begin{align}
        \nonumber (q_1,q_2,q_3)=\mathop{argmin}_{q_1,q_2,q_3} \prod_i (1+e^{-\alpha q_i}) (e^{gy\cdot 1}+e^{gy\cdot e^{i\frac{\pi}{4}}+\alpha q_3}+e^{ gy\cdot e^{i\frac{\pi}{2}}+\alpha(q_2+q_3)}\\ \nonumber +e^{gy\cdot e^{i\frac{3\pi}{4}}+\alpha q_2}+e^{ gy\cdot e^{i\pi}+\alpha (q_1+q_2)}+e^{gy\cdot e^{i\frac{5\pi}{4}}+\alpha (q_1+q_2+q_3)}\\ +e^{ gy\cdot e^{i\frac{3\pi}{2}}+\alpha (q_1+q_3)}+e^ {gy\cdot e^{i\frac{7\pi}{4}}+\alpha q_1})
        \label{smallgcr}
        \end{align} 
        
        

\begin{figure}[ht!]
	\centering
	\includegraphics[width=80mm]{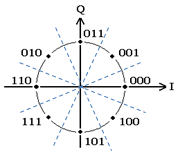}
	\caption{The 8-region devision of the channel output space for 8PSK demodulation.}
	\label{8pd}
\end{figure}

To simplify our discussion, we divide the channel output space into 8 symmetric regions (see Fig. \ref{8pd}) based on the complex phase of the channel output $y$. We shall only derive the solution to (\ref{smallgcr}) when the channel output phase belongs to $[-\frac{\pi}{8},\frac{\pi}{8}]$, and the rest part of the solution can be easily obtained due to the symmetry.

When the channel output belongs to the region specified above, we have the following inequalities
       \begin{align}
       y\cdot 1\geq  y\cdot e^{\pm i\frac{\pi}{4}}\geq  y\cdot e^{ \pm i\frac{\pi}{2}}\geq  y\cdot e^{\pm i\frac{3\pi}{4}}\geq  y\cdot e^{ i\pi}
       \label{fct}
       \end{align} 
       
Then the target function defined in (\ref{smallgcr}) is lower bounded by $e^ {y\cdot 1}$, and this exponent can be achieved when $e^{\alpha q_i}\gg1$ and the following inequalities hold. 
          \begin{align}
          \left \{\begin{array}{@{}lll@{}}
          & y\cdot 1\geq&   y\cdot e^{i\frac{\pi}{4}}+\frac{\alpha}{g} q_3 \\
          & y\cdot 1\geq&   y\cdot e^{i\frac{\pi}{2}}+\frac{\alpha}{g} (q_2+q_3)\\
          & y\cdot 1\geq&  y\cdot e^{i\frac{3\pi}{4}}+\frac{\alpha}{g} q_2\\
          & y\cdot 1\geq&  y\cdot e^{i\pi}+\frac{\alpha}{g} (q_1+q_2)\\
          & y\cdot 1\geq&  y\cdot e^{i\frac{5\pi}{4}}+\frac{\alpha}{g} (q_1+q_2+q_3)\\
          & y\cdot 1\geq&  y\cdot e^{i\frac{3\pi}{2}}+\frac{\alpha}{g} (q_1+q_3)\\
          & y\cdot 1\geq& y\cdot e^{i\frac{7\pi}{4}}+\frac{\alpha}{g} q_1 
          \end{array}
          \right.
          \end{align} 
     

Applying (\ref{fct}) to the above inequalities, we can derive a simplified version of the constraints on the optimal bit metrics.
          \begin{align}
          \label{con}
          \left \{\begin{array}{@{}lll@{}}
          & q_i \geq 0\\
          & y\cdot 1\geq&   y\cdot e^{i\frac{\pi}{4}}+\frac{\alpha}{g} q_3 \\
          & y\cdot 1\geq&   y\cdot e^{i\frac{\pi}{2}}+\frac{\alpha}{g} (q_2+q_3)\\
          & y\cdot 1\geq& y\cdot e^{i\frac{7\pi}{4}}+\frac{\alpha}{g} q_1 
          \end{array}
          \right.
          \end{align}

          Now we start solving for the optimal values of the decoding metrics. Taking the partial derivative of the objective function in (\ref{smallgcr}) with respect to the bit metrics, we obtain the equations that restrict its solution.
          \begin{align}
          2\alpha q_1&=log \frac{e^{gy\cdot 1}+e^{gy\cdot e^{i\frac{\pi}{4}}+\alpha q_3}+e^{ gy\cdot e^{i\frac{\pi}{2}}+\alpha(q_2+q_3)}+e^{gy\cdot e^{i\frac{3\pi}{4}}+\alpha q_2}}{e^{ gy\cdot e^{i\pi}+\alpha q_2}+e^{gy\cdot e^{i\frac{5\pi}{4}}+\alpha (q_2+q_3)}+e^{ gy\cdot e^{i\frac{3\pi}{2}}+\alpha q_3}+e^ {gy\cdot e^{i\frac{7\pi}{4}}}}\\
          2\alpha q_2&=log \frac{e^{gy\cdot 1}+e^{gy\cdot e^{i\frac{\pi}{4}}+\alpha q_3}+e^{ gy\cdot e^{i\frac{3\pi}{2}}+\alpha (q_1+q_3)}+e^ {gy\cdot e^{i\frac{7\pi}{4}}+\alpha q_1}}{e^{ gy\cdot e^{i\frac{\pi}{2}}+\alpha q_3}+e^{gy\cdot e^{i\frac{3\pi}{4}}}+e^{ gy\cdot e^{i\pi}+\alpha q_1}+e^{gy\cdot e^{i\frac{5\pi}{4}}+\alpha (q_1+q_3)}}\\
          2\alpha q_3&=log \frac{e^{gy\cdot 1}+e^{gy\cdot e^{i\frac{3\pi}{4}}+\alpha q_2}+e^{ gy\cdot e^{i\pi}+\alpha (q_1+q_2)}+e^ {gy\cdot e^{i\frac{7\pi}{4}}+\alpha q_1}}{e^{gy\cdot e^{i\frac{\pi}{4}}}+e^{ gy\cdot e^{i\frac{\pi}{2}}+\alpha q_2}+e^{gy\cdot e^{i\frac{5\pi}{4}}+\alpha (q_1+q_2)}+e^{ gy\cdot e^{i\frac{3\pi}{2}}+\alpha q_1}}
          \end{align} 
          
          Utilizing in inequalities in (\ref{con}) and the max-log approximation, the above equations can be simplified into
          \begin{align}
          2\alpha q_1&=gy\cdot 1-\max\{{ gy\cdot e^{i\pi}+\alpha q_2},\ { gy\cdot e^{i\frac{3\pi}{2}}+\alpha q_3},\ {gy\cdot e^{i\frac{7\pi}{4}}}\}\\
          2\alpha q_2&=gy\cdot 1-{ gy\cdot e^{i\frac{\pi}{2}}-\alpha q_3}\\
          2\alpha q_3&=gy\cdot 1-\max\{{gy\cdot e^{i\frac{\pi}{4}}},\ { gy\cdot e^{i\frac{\pi}{2}}+\alpha q_2}\}
          \end{align} 
             
        Solving the above equations, we derive the formulas for the optimal solution, which can also be expressed using their corresponding log-likelihood ratios.
        
        \begin{align}
        q_1&=\frac{gy\cdot (1-e^{i\frac{7\pi}{4}})}{2\alpha} = \frac{LLR_1}{2\alpha}\\
        q_2&=\frac{gy\cdot (1+e^{i\frac{\pi}{4}}-2e^{i\frac{\pi}{2}})}{4\alpha} = \frac{LLR_2-\frac{1}{2} LLR_3}{2\alpha}\\
        q_3&=\frac{gy\cdot (1-e^{i\frac{\pi}{4}})}{2\alpha} = \frac{LLR_3}{2\alpha}
        \end{align} 
        
        Since scaling the quantization metrics does not affect the decoding result, the parameter $\alpha$ can be an arbitrary positive value.
\begin{singlespace}
\bibliography{main}

\newcommand{\noopsort}[1]{} \newcommand{\printfirst}[2]{#1}
  \newcommand{\singleletter}[1]{#1} \newcommand{\switchargs}[2]{#2#1}
\begin{thebibliography}{10}

\bibitem{abbe2007finding}
Emmanuel Abbe, Muriel M{\'e}dard, Sean Meyn, and Lizhong Zheng.
\newblock Finding the best mismatched detector for channel coding and
  hypothesis testing.
\newblock In {\em Information Theory and Applications Workshop, 2007}, pages
  284--288. IEEE, 2007.

\bibitem{ahlswede1996erasure}
Rudolf Ahlswede, Ning Cai, and Zhen ZHang.
\newblock Erasure, list, and detection zero-error capacities for low noise and
  a relation to identification.
\newblock {\em Information Theory, IEEE Transactions on}, 42(1):55--62, 1996.

\bibitem{balakirsky1995converse}
Vladimir~B Balakirsky.
\newblock A converse coding theorem for mismatched decoding at the output of
  binary-input memoryless channels.

\bibitem{barre2013polar}
Anthony Barr{\'e}, Emmanuel Boutillon, Neysser Bias, and Daniel Diaz.
\newblock A polar-based demapper of 8psk demodulation for dvb-s2 systems.
\newblock In {\em Signal Processing Systems (SiPS), 2013 IEEE Workshop on},
  pages 13--17. IEEE, 2013.

\bibitem{biederman1981decoding}
Leon Biederman, J~Omura, and P~Jain.
\newblock Decoding with approximate channel statistics for bandlimited
  nonlinear satellite channels.
\newblock {\em Information Theory, IEEE Transactions on}, 27(6):697--707, 1981.

\bibitem{binshtok1999integer}
Nir Binshtok and Shlomo Shamai.
\newblock Integer metrics for binary input symmetric output memoryless
  channels.
\newblock {\em Communications, IEEE Transactions on}, 47(11):1636--1645, 1999.

\bibitem{borjesson1979simple}
Per~Ola B{\"o}rjesson and Carl-Erik~W Sundberg.
\newblock Simple approximations of the error function q (x) for cummunications
  applications.
\newblock {\em IEEE Transactions on Communications}, 27(3):639--643, 1979.

\bibitem{chandrasekaran2011capacity}
Suresh Chandrasekaran, Saif~K Mohammed, and Ananthanarayanan Chockalingam.
\newblock On the capacity of quantized gaussian mac channels with finite input
  alphabet.
\newblock In {\em Communications (ICC), 2011 IEEE International Conference on},
  pages 1--5. IEEE, 2011.

\bibitem{cheng2009structured}
Michael~K Cheng, Dariush Divsalar, and Stephanie Duy.
\newblock Structured low-density parity-check codes with bandwidth efficient
  modulation.
\newblock In {\em SPIE Defense, Security, and Sensing}, pages 73490C--73490C.
  International Society for Optics and Photonics, 2009.

\bibitem{danieli2010maximum}
Matteo Danieli, S{\o}ren Forchhammer, Jakob~Dahl Andersen, Lars~PB Christensen,
  and S{\o}ren~Skovgaard Christensen.
\newblock Maximum mutual information vector quantization of log-likelihood
  ratios for memory efficient harq implementations.
\newblock In {\em Data Compression Conference (DCC), 2010}, pages 30--39. IEEE,
  2010.

\bibitem{hui1983fundamental}
Joseph Yu~Ngai Hui.
\newblock Fundamental issues of multiple accessing.
\newblock 1983.

\bibitem{ibragimov1972weak}
Il'dar~Abdullovich Ibragimov and Rafail~Zalmanovich Khas'~minskii.
\newblock Weak signal transmission in a memoryless channel.
\newblock {\em Problemy Peredachi Informatsii}, 8(4):28--39, 1972.

\bibitem{jalden2010generalized}
Joakim Jald{\'e}n, Peter Fertl, and Gerald Matz.
\newblock On the generalized mutual information of bicm systems with
  approximate demodulation.
\newblock In {\em Information Theory Workshop (ITW), 2010 IEEE}, pages 1--5.
  IEEE, 2010.

\bibitem{liveris2003quantization}
Angelos~D Liveris and Costas~N Georghiades.
\newblock On quantization of low-density parity-check coded channel
  measurements.
\newblock In {\em Global Telecommunications Conference, 2003. GLOBECOM'03.
  IEEE}, volume~3, pages 1649--1653. IEEE, 2003.

\bibitem{martinez2009bit}
Alfonso Martinez, Albert Guill{\'e}n~i F{\`a}bregas, Giuseppe Caire, and
  Frans~MJ Willems.
\newblock Bit-interleaved coded modulation revisited: A mismatched decoding
  perspective.
\newblock {\em Information Theory, IEEE Transactions on}, 55(6):2756--2765,
  2009.

\bibitem{max1960quantizing}
Joel Max.
\newblock Quantizing for minimum distortion.
\newblock {\em Information Theory, IRE Transactions on}, 6(1):7--12, 1960.

\bibitem{merhav1994information}
Neri Merhav, Gideon Kaplan, Amos Lapidoth, and S~Shamai~Shitz.
\newblock On information rates for mismatched decoders.
\newblock {\em Information Theory, IEEE Transactions on}, 40(6):1953--1967,
  1994.

\bibitem{nguyen2011bit}
Trung~Thanh Nguyen and Lutz Lampe.
\newblock Bit-interleaved coded modulation with mismatched decoding metrics.
\newblock {\em Communications, IEEE Transactions on}, 59(2):437--447, 2011.

\bibitem{novak2009quantization}
Clemens Novak, Peter Fertl, and Gerald Matz.
\newblock Quantization for soft-output demodulators in bit-interleaved coded
  modulation systems.
\newblock In {\em Information Theory, 2009. ISIT 2009. IEEE International
  Symposium on}, pages 1070--1074. IEEE, 2009.

\bibitem{panter1951quantization}
PF~Panter and W~Dite.
\newblock Quantization distortion in pulse-count modulation with nonuniform
  spacing of levels.
\newblock {\em Proceedings of the IRE}, 39(1):44--48, 1951.

\bibitem{park2008low}
Jang~Woong Park, Myung~Hoon Sunwoo, Pan~Soo Kim, and Dae-Ig Chang.
\newblock Low complexity soft-decision demapper for high order modulation of
  dvb-s2 system.
\newblock In {\em SoC Design Conference, 2008. ISOCC'08. International},
  volume~2, pages II--37. IEEE, 2008.

\bibitem{prelov1993asymptotic}
Vyacheslav~V Prelov and Edward~C van~der Meulen.
\newblock An asymptotic expression for the information and capacity of a
  multidimensional channel with weak input signals.
\newblock {\em Information Theory, IEEE Transactions on}, 39(5):1728--1735,
  1993.

\bibitem{rave2009quantization}
Wolfgang Rave.
\newblock Quantization of log-likelihood ratios to maximize mutual information.
\newblock {\em Signal Processing Letters, IEEE}, 16(4):283--286, 2009.

\bibitem{salz1995decoding}
Jack Salz and Ephraim Zehavi.
\newblock Decoding under integer metrics constraints.
\newblock {\em Communications, IEEE Transactions on}, 43(2/3/4):307--317, 1995.

\bibitem{shahid2008distributed}
Khattak Shahid, Rave Wolfgang, and Fettweis Gerhard.
\newblock Distributed iterative multiuser detection through base station
  cooperation.
\newblock {\em EURASIP Journal on Wireless Communications and Networking},
  2008, 2008.

\bibitem{shannon2001mathematical}
Claude~Elwood Shannon.
\newblock A mathematical theory of communication.
\newblock {\em ACM SIGMOBILE Mobile Computing and Communications Review},
  5(1):3--55, 2001.

\bibitem{somekh2014general}
Anelia Somekh-Baruch.
\newblock A general formula for the mismatch capacity.
\newblock In {\em Information Theory (ISIT), 2014 IEEE International Symposium
  on}, pages 3067--3071. IEEE, 2014.

\bibitem{zehavi19928}
Ephraim Zehavi.
\newblock 8-psk trellis codes for a rayleigh channel.
\newblock {\em Communications, IEEE Transactions on}, 40(5):873--884, 1992.

\end{thebibliography}
\bibliographystyle{plain}

\end{singlespace}

\end{document}